\documentclass[]{raa}            % referee version: for submission

\usepackage{graphicx,times}             %for PS/EPS graphics inclusion, new
\usepackage{natbib}
\usepackage{amssymb,amsmath}
\bibpunct{(}{)}{;}{a}{}{,}

\usepackage[pagebackref=true]{hyperref}

\usepackage[bottom = 0.5in]{geometry}

\begin{document}

  \title{AT2021acak: a candidate tidal disruption event found in the Zwicky Transient Facility survey
}
%%   \subtitle{A candidate tidal disruption event}

   \volnopage{Vol.0 (20xx) No.0, 000--000}      %%preserved for Editor. DOn't remove!
   \setcounter{page}{1}          %%starting page, preserved for Editor. DOn't remove!

   \author{Jie Li %(周爱英) %% Put your Chinese name in "( )" if you like. Note to open line 11 "\usepackage[UTF8]{ctex}"
      \inst{1}
   \and Zhong-Xiang Wang
      \inst{1,2,3}
   \and Dong Zheng
      \inst{1}
   \and Ju-Jia Zhang
   	\inst{4,5,6}
   \and Li-Tao Zhu
        \inst{1}
   \and Zhang-Yi Chen
   	\inst{1}
   }

   \institute{Department of Astronomy, School of Physics and Astronomy, Yunnan University, Kunming 650091, China; {\it wangzx20@ynu.edu.cn}\\
        \and
Shanghai Astronomical Observatory, Chinese Academy of Sciences, 80 Nandan Road, Shanghai 200030, China\\
	\and 
Purple Mountain Observatory, Chinese Academy of Sciences, Nanjing 210034, China\\
        \and
	Yunnan Observatories, Chinese Academy of Sciences, Kunming 650216, China\\
	\and
	Key Laboratory for the Structure and Evolution of Celestial Objects, Chinese Academy of Sciences, Kunming 650216, China\\
	\and
	Center for Astronomical Mega-Science, Chinese Academy of Sciences, 20A Datun Road, Chaoyang District, Beijing 100012, China\\
\vs\no
   {\small Received 20xx month day; accepted 20xx month day}}

\abstract{We report a candidate tidal disruption event (TDE) found in the Zwicky
Transient Facility (ZTF) survey data. This candidate, with its 
transient name AT2021acak, showed brightness increases of 
$\sim$1\,mag around MJD~59500
and subsequent power-law--like brightness declines. We have conducted 
multiple optical spectroscopic observations with the 2.4-m Lijiang telescope
and one observation at X-ray and ultraviolet (UV) bands with 
the {\it Neil Gehrels Swift Observatory (Swift)}. The optical spectra of the 
source show broad H and He emission line and Fe emission features.
Possible 0.3--2\,keV X-ray and bright UV emission of the source 
was detected.  We analyze the declines of the optical
light curves, the emission features of the optical spectra, and the constructed
broad-band UV and optical spectrum. The properties derived from the
analyses are consistent with those of reported (candidate) TDEs, and 
in particular very similar to those of ASASSN-18jd. 
The identification is complicated by the host being likely an AGN, and thus
further observations of the event and quiescent host are required in order to
have a clear understanding of the nature of this transient event.
\keywords{transients: tidal disruption events --- galaxies: nuclei --- quasars: emission}
}

   \authorrunning{Li et al.}            %author_head in even pages
   \titlerunning{A candidate tidal disruption event}  % title_head in odd pages

   \maketitle
%% The author head (on even pages) and the title head (on odd pages) will be
%% automatically extracted from \author{} and \title{}. Whenever the title is too long,
%% you will be asked to supply a shorter one by inserting either \authorrunning{} or
%% \titlerunning{} before \maketitle. Anyway, you can specify your own heads.
%%
%%
%% Note: In the following text body of your manuscript, please note several differences from
%%       other major journals:
%% (1) \subsection{Please Capitalize the First Letter of Each Notional Word in Subsection Title}
%% (2) Please Capitalize the First Letter of Each Notional Word in all tables' captions

%
%________________________________________________ sections below
%
\section{Introduction}         %% first-level sections will be auto-capitalized
\label{sect:intr}

When a star moves too closely to the central massive black hole (MBH) of
a galaxy, for example
$\sim$1\,AU for a solar-mass star to a 10$^7\,M_{\odot}$ MBH, it will 
be under tidal forces exerted by the MBH that are stronger than its own 
gravitational force, and thus be disrupted. Approximately half of the star's 
mass will be
bounded by the gravitational potential of and subsequently accreted by
the MBH.  During this so-called tidal disruption event (TDE), the accretion
will power transient radiation from the MBH that is otherwise often in
a non-active state. While the TDEs have been long proposed and widely explored
theoretically 
(e.g., \citealt{hil75,ree88,ek89,sq09,lr11,gmr14,koc16,dai+18,mgr19,cn19}), 
more and more observational discoveries of them have been reported 
in recent years (e.g., \citealt{van+21,gez21})
enabled by the fast-growing capability of the transient surveys.
Studies of TDEs can probe properties of dormant MBHs in galaxies, 
transient accretion processes, and stellar dynamics close to the center
of galaxies (see \citealt{gez21} and references therein).

Main observational discoveries of TDEs started with the ROSAT all-sky survey 
at X-rays
\citep{don+02}, and continued with the transient surveys at optical
wavelengths such as the Palomar Transient Factory (e.g., \citealt{arc+14})
and All-Sky Automated Survey for Supernovae (e.g., \citealt{hol+14}).
As the Zwicky Transient Facility (ZTF; \citealt{bel+19}) represents 
the latest, more powerful
optical surveys, with respect to its large field of view (47\,deg$^2$) and
high cadence ($\sim$2\,day), it can serve as a powerful facility for finding
TDEs (and other similar transient events). Indeed, \citet{van+21} reported
17 new TDEs recently detected in the ZTF survey. These TDEs, plus
the previously reported ones, show a variety of features that not only reveal 
the accretion-related physical processes but also present different challenges
to theoretical studies, thus helping improve our full understanding of the TDE 
phenomenon \citep{gez21}.

As we have been exploring the ZTF data, one source (named AT2021acak
in the transient name server\footnote{https://www.wis-tns.org/object/2021acak})
was found to have the light curves resembling those of the TDEs. 
Simultaneous infrared brightening of the source was also found in the NEOWISE 
Post-Cryo survey data obtained with the Wide-field Infrared Survey 
(WISE; \citealt{wri+10}). We thus arranged 
optical spectroscopic observations with the 2.4-m Lijiang telescope and 
a Target of Opportunity (ToO) observation at X-ray
and ultraviolet (UV) bands with {\it the Neil Gehrels Swift Observatory
(Swift)} \citep{geh+04}. In this paper, we report the results from 
the observations for this transient event case. 

AT2021acak arises from a source with a position of R.A.=$\rm10^h34^m47^s.99$, 
Decl.=$+15^{\circ}29'22''.42$ (J2000.0). Hereafter we use name J1034 for 
this source. 
As there are previous observational data for the source obtained 
from different surveys, we first collected the information and studied
its likely
source type. A brief description of the data and our source-type analysis
is presented in Section~\ref{subsec:src}.
We then describe the data that were taken during the time period just
before and over the transient event of the
source. The data include those from the ZTF and WISE surveys
and our observations as well. The data description and reduction processes
are given in Section~\ref{sect:data}.
The related analyses and results are presented in Section~\ref{sect:res}. 
We discuss the results in Section~\ref{sect:dis}, which
points out this transient case as a candidate TDE.
In this work
we used the latest cosmological parameters determined from the Planck 
mission \citep{planck18}, 
$H_0\simeq 67.4$\,km\,s$^{-1}$\,Mpc$^{-1}$ and $\Omega_m=0.315$.
\begin{figure}
   \centering
   \includegraphics[width=0.7\textwidth, angle=0]{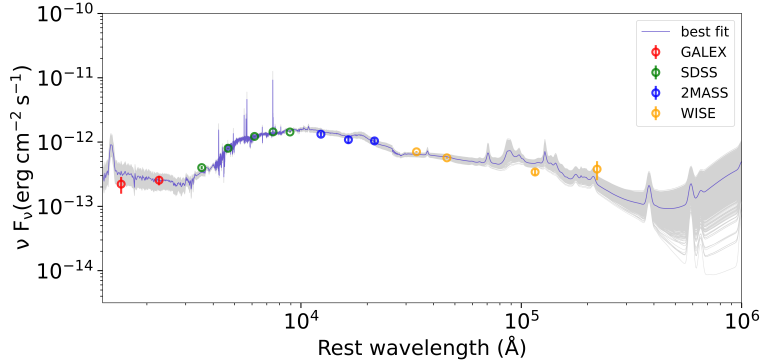}
	\caption{Broad-band SED of J1034+1529 in its presumably quiescent state.
	It can be fit with
	AGN type spectra, which are indicated by the gray line region and
	were generated from the code {\tt PROSPECTOR}. }
   \label{fig:sed}
\end{figure}

\subsection{Source J1034+1529}
\label{subsec:src}

The source J1034 has been detected in different surveys, which include
the Galaxy Evolution Explorer (GALEX; \citealt{galex11}), the Sloan Digital 
Sky Survey (SDSS; \citealt{sdss09}), the Two Micron All Sky Survey 
(2MASS; \citealt{2mass03}), and the WISE \citep{wise12}.
Its infrared colors were
$J-Ks=1.31$ and $Ks-W3=4.01$ or mid-infrared (MIR) colors were
$W1-W2=0.78$ and $W2-W3=2.19$. The both suggest a quasar
(e.g., \citealt{wri+10,tw13}). Indeed, J1034 is included as a quasar
in the Million Quasar catalog (version 7.2; \citealt{fle21}).

We constructed its broad-band spectral energy distribution (SED;
see Figure~\ref{fig:sed}) based
the flux measurements mentioned above, which were approximately obtained before
2012. While the measurements were not simultaneous and the SED might not be
a true `quiescent' one, we used the code {\tt PROSPECTOR} \citep{joh+21} to
fit the SED. The result is shown in Figure~\ref{fig:sed}, for which 
the parameter $f_{\rm agn}\simeq 0.27$ (see \citealt{lej+18} for details).
This parameter value suggests a significant AGN contribution in the emission
of J1034.

We note that the source also showed flux variations at optical bands. 
Revealed by the optical light curves obtained with the Asteroid 
Terrestrial-impact Last Alert System (ATLAS; \citealt{ton+18}), a $\sim$100\,d
flare-like event around MJD~57800 is seen (Figure~\ref{fig:lc}). The
event lasted relatively short and then the source went back to its
previous quiescent magnitude level. Thus this variation event likely is another
piece of evidence, in addition to the above, that points to an AGN for J1034.

\section{Data and Observations}
\label{sect:data}

\subsection{Light Curve data}

\begin{figure}
   \centering
   \includegraphics[width=0.7\textwidth, angle=0]{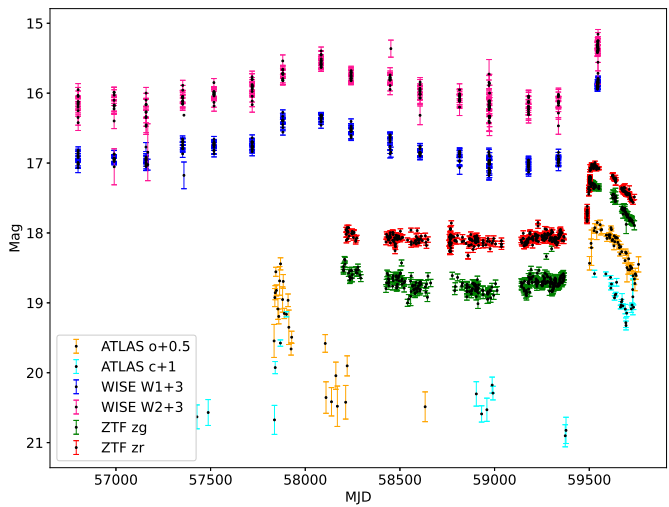}
   \caption{Optical and MIR light curves of J1034+1529, the latter of
	which are down-shifted by 3 mag to be shown in one panel. 
	The ATLAS light curves, which were constructed by excluding
	data points with either magnitudes $\geq$20 \citep{ton+18}
	or uncertainties $>$0.3\,mag, show a flare-like event around MJD~57800.
	Around MJD~59500, there is a brightness jump seen in the both ATLAS
	and ZTF optical and WISE MIR bands.}
   \label{fig:lc}
\end{figure}

As we explored the ZTF light curve data with the Automatic Learning for the
Rapid Classification of Events (ALeRCE; \citealt{for+21,san+21}), we noted
this source's recent flaring-like activity starting from $\sim$MJD~59500:
after the sudden brightening,
the light curves have been going through a power-law decline.
We downloaded the $zg$ and $zr$ band light curve data from the ZTF public 
Data Release 11
(released on 2022 May 9). For keeping the cleanness and goodness of 
the data, we required catflags=0 and $chi<4$ when querying the ZTF data.
Also in order to have as many data as possible, we included those data
points available through ALeRCE\footnote{https://alerce.online/}.
The $zg$ and $zr$ light curves are shown in Figure~\ref{fig:lc}. 
The ATLAS light curves show a similar pattern. However because
their bands are not regular ones and the limiting magnitudes are only slightly
larger than 19 \citep{ton+18}, the latter resulting in only a few data points
before the brightening, we did not include the ATLAS light curves in the further
analysis.

We also checked the source's MIR W1 (3.4\,$\mu$m) and W2 ($4.6\,\mu$m)
light curve data, and downloaded them from the WISE NEOWISE-R Single-exposure 
Source database. In Figure~\ref{fig:lc}, the MIR light curves are shown.
As can be seen, a related brightening also occurred at the MIR
wavelengths, although the data were basically from one-epoch of observations.
The average magnitudes (in MJD~59000--59500) of the source before 
the brightening are W1=14.00 and W2=13.17, and at the brightening (average
time MJD~59545) are W1=12.87 and W2=12.35. Therefore the source brightened
by 1.1 and 0.8 mag at W1 and W2 bands respectively.

\subsection{Spectroscopic observations}

Five spectroscopic observations of J1034 were conducted with the 2.4-m Lijiang
telescope. The instrument used was YFOSC \citep{wan+19}. Its detector 
is a $2048\times 4608$ pixel$^2$ back-illuminated CCD and 
the pixel scale is 0.283$''$\,pixel$^{-1}$. In the observations, 
a G15 (or G3) grism was chosen, providing the wavelength coverage of 
4100--9800\,\AA\ (or 3400--9100\,\AA) and the spectral dispersion
of 3.9\,\AA\,pixel$^{-1}$ (or 2.9\,\AA\,pixel$^{-1}$). 
A long slit was used, and its width was set to be
either 5.0$''$ or 2.5$''$ (see Table~\ref{tab:obs}). 
In each observation, 
we also took the spectrum of a He-Ne lamp for wavelength calibration and
that
of the spectrophotometric standard BD+33d2642 for flux calibration,
while with the same instrument setup as that for the target.
The information for 
the observations, including the exposure times, are summarized
in Table~\ref{tab:obs}. 

The spectrum images were routinely processed using the IRAF tasks, which
included bias-subtraction and flat-fielding. Spectrum extraction and calibration
were conducted using the standard routines in IRAF as well.

\begin{table}
\bc
\begin{minipage}[]{100mm}
\caption{Information for spectroscopic observations of J1034\label{tab:obs}}\end{minipage}
\setlength{\tabcolsep}{7pt}
\small
\begin{tabular}{lcccc}
\hline\noalign{\smallskip}
        Date    &  Grism     & Slit width     & Seeing         & Exposure   \\
        (UT)    &            &  ($''$)   & ($''$)   & (s)              \\
\hline\noalign{\smallskip}
        2022-04-23 & G15      & 5.0           & 1.8           & 900          \\ 
        2022-04-23  & G15     & 5.0           & 1.8           & 900          \\ 
        2022-04-24  & G3      & 2.5           & 2.1           & 900          \\ 
        2022-04-25  & G3      & 2.5           & 2.1           & 1200         \\ 
        2022-06-03  & G3      & 2.5           & 1.7           & 1350         \\ 
\noalign{\smallskip}\hline
\end{tabular}
\ec
\end{table}
      
\subsection{{\it Swift} Observation}

We requested a ToO observation of the source with 
{\it Swift}. The observation (obsid 00015141001) was conducted on 2022 
April 28. The photon counting mode and image mode were respectively used in
the X-ray Telescope (XRT) and the Ultraviolet/Optical Telescope
(UVOT) exposures. The lengths of the exposures were approximately 1500-s long,
and the filter UVW2 (central wavelength 
1928\,\AA; \citealt{poo+08}) was used in the UVOT exposure.

We used the online tools\footnote{https://www.swift.ac.uk/user\_objects/}
for the XRT data analysis and source detection \citep{eva+20}. No detection
of the source was found, while we tested different energy bands such as
0.3--1\,keV, 1--2\,keV, and 2--10\,keV. However when we used the light curve 
tool \citep{eva+07,eva+09}, a count rate of 
($4.6\pm2.1)\times 10^{-3}$\,cts\,s$^{-1}$ in 0.3--2\,keV was derived, which
was indicated to be higher than that of the background by a 3$\sigma$ 
confidence level.  We checked the individual photons but could not determine 
if this marginal detection was 
real or not. 
We considered this count rate as an upper limit (but discuss 
the detection possibility in Section~\ref{sect:dis}).
In 2--10\,keV,
an upper limit of $5.3\times 10^{-3}$\,cts\,s$^{-1}$ (3$\sigma$) was obtained.

In order to convert the count rate to flux, we used the tool PIMMS. We tested
blackbody and power-law models, where the Galactic hydrogen column density
$3.1 \times 10^{20}$\,cm$^{-2}$ \citep{hi4pi+16} was assumed. 
The resulting unabsorbed flux upper limits in 0.3--2\,keV were in a range of 
1.1--1.6$\times 10^{-13}$\,erg\,cm$^{-2}$\,s$^{-1}$, depending on the model
assumed. We adopted the middle value,
$1.4\times 10^{-13}$\,erg\,cm$^{-2}$\,s$^{-1}$, as the flux upper limit
for the source.

The source was detected in the UVOT exposure. We ran the tool {\tt uvotsource} 
and obtained a magnitude of 16.71$\pm$0.05 for the source, where a circular
source region with radius 5$''$ and a circular background region with 
radius 15$''$ were used. The magnitude corresponds to a flux of 
2.14$\pm0.10\times 10^{-12}$\,erg\,cm$^{-2}$\,s$^{-1}$.

\begin{figure}
   \centering
   \includegraphics[width=0.52\textwidth, angle=0]{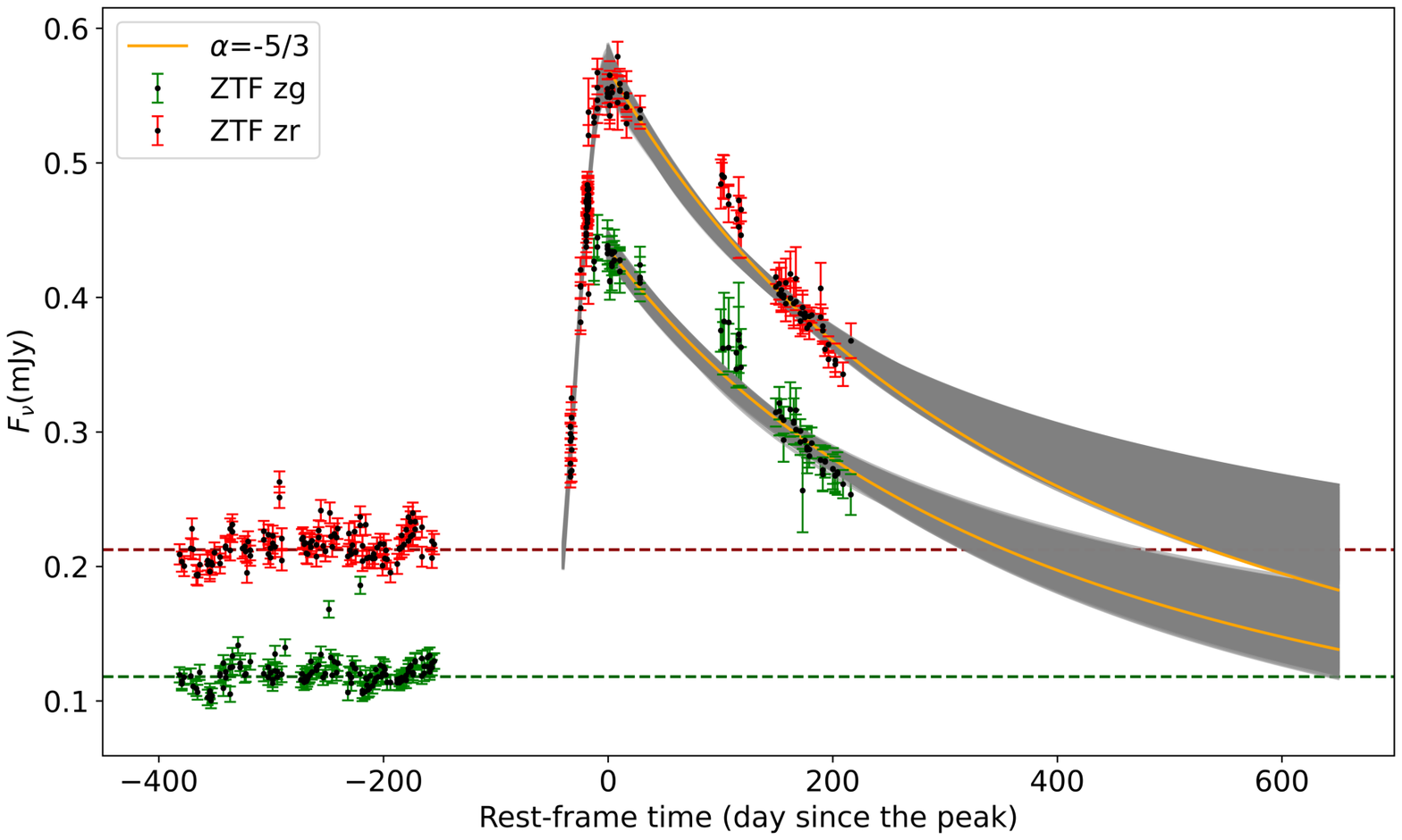}
   \includegraphics[width=0.43\textwidth, angle=0]{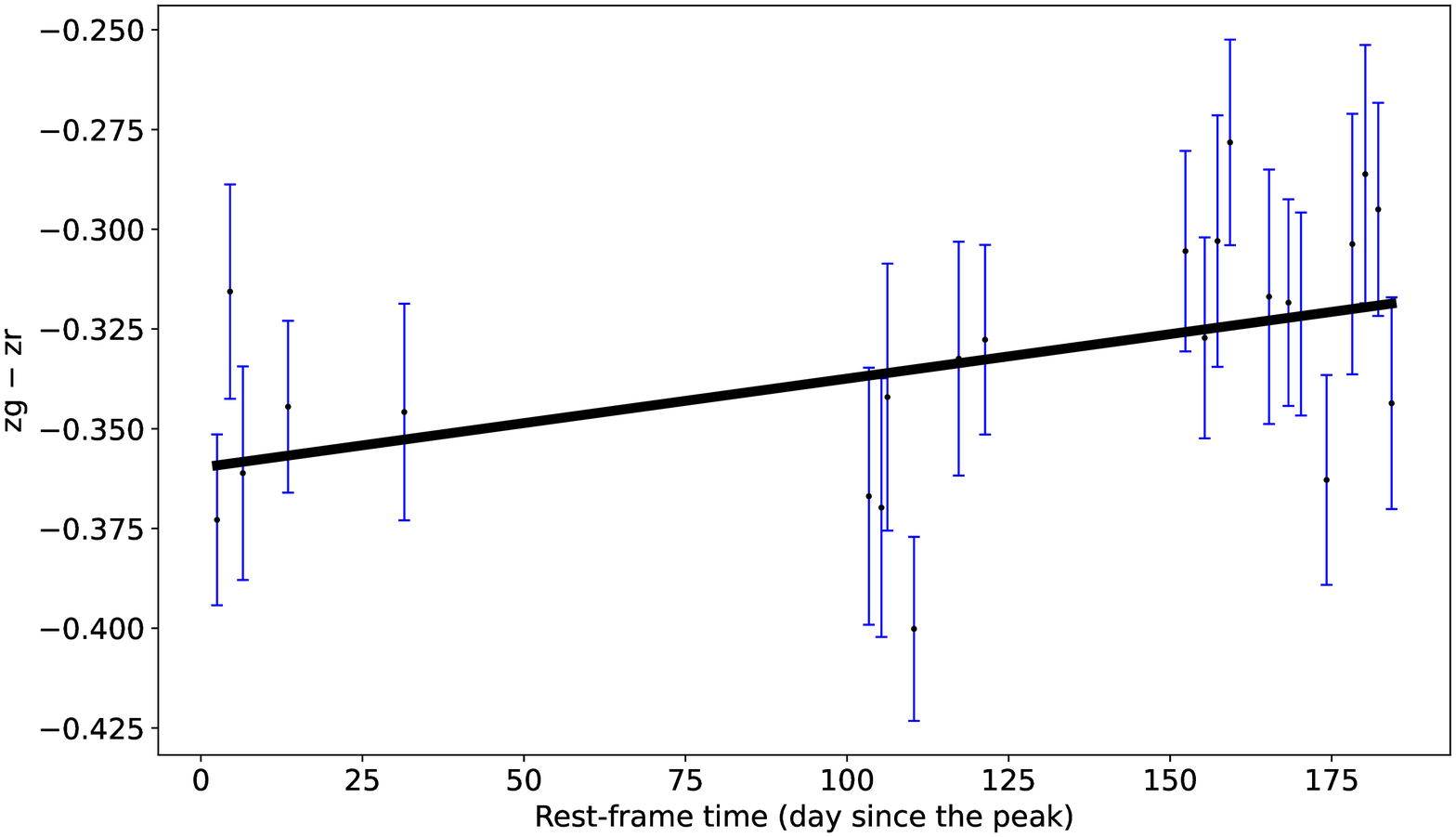}
	\caption{{\it Left} panel: Optical light curve fitting with a typical 
	TDE model (gray regions). The flux density declines since the peak 
	as a function of time $t^{-5/3}$ are marked with orange curves. 
	The average fluxes at $zg$ and $zr$ bands before the flare are 
	marked with the dashed lines. 
	{\it Right} panel: $zg-zr$ colors
	during the flare, for which a rate of 
	2.2$\pm0.8\times 10^{-4}$\,day$^{-1}$ (black line) is derived.}
   \label{fig:lcfit}
\end{figure}

\section{Results}
\label{sect:res}

\subsection{Optical light curves}
The detailed optical light curves before (in quiescence) and covering 
the flare event are shown in the left panel of
Figure~\ref{fig:lcfit}. The average fluxes in quiescence (shown in 
Figure~\ref{fig:lcfit}) are 
0.118$\pm0.012$ and 0.213$\pm0.014$\,mJy in
$zg$ and $zr$ band respectively, where the uncertainties were estimated
as the standard deviations of each quiescent light-curve part.
We studied the light curves by fitting them with a model typically 
considered for TDEs.
The model (e.g., \citealt{van+21}) is given as 1)
$F(t)=F_p\exp[-(t-t_p)^2/2\sigma_t^2$] when $t\leq t_p$, and 2)
$F(t)=F_p[(t-t_p+t_0)/t_0]^{\alpha}$ when $t\geq t_p$, where
$t$ is time and the flux $F(t)$ has a peak $F_p$ at time $t_p$. In this model,
the first function
is a Gauss with standard deviation $\sigma_t$ and the second is a power law
with index $\alpha$ and normalization $t_0$.

We used the Markov Chain Monte Carlo (MCMC) code {\tt emcee} \citep{for+13}
for the fitting, in which the times of the data points were shifted to the rest
frame with $z=0.136$ (see Section~\ref{subsec:spec}). As the flux rise of 
the event was caught in $zr$ band,
this part was fitted with the Gaussian function and $t_p$ and $\sigma_t$
were determined to be MJD~59522 and 27\,day. The event would have started
from MJD~59489. The peak fluxes were 0.44 and 0.58\,mJy in $zg$ and $zr$ band
respectively.
We then fixed $t_p=59522$ and fit the decline parts of the light curves 
with the power-law function. The function can approximately describe the light
curves, but a few data points in the middle appear to deviate away 
from a power law (Figure~\ref{fig:lcfit}). The index ranges (1$\sigma$) 
found from the
fitting were ($-2.08$, $-0.64$) for the $zg$ band and ($-1.26$, $-0.61$) for 
the $zr$ band. Both the ranges are large, probably due to the sparseness
of the data points, and the second one is flatter than the typical $-5/3$
decline \citep{cn19}. We suspect that
there could be a break in the decline parts; for example in $zg$ band, the
light curve could be considered to be
relatively flat from the peak to the middle (with $\alpha \sim -0.6$) and turn
steep in the latter part (with $\alpha \sim -1.2$).
\begin{figure}
   \centering
   \includegraphics[width=0.7\textwidth, angle=0]{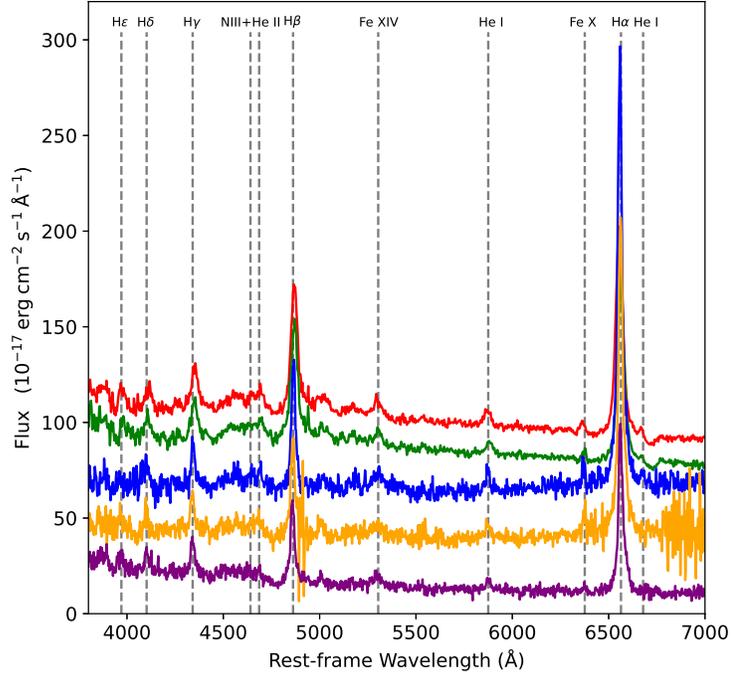}
	\caption{Optical spectra of J1034, which follow the sequence 
	given in Table~\ref{tab:obs} from the above to bottom.
	For clarity, they are slightly shifted in the vertical direction.
	A few emission lines detected
	in the spectra are marked, which include Balmer lines, He~I $\lambda$5875, and He~II $\lambda$4686. Also weak [Fe~XIV] $\lambda$5304 and [Fe~X] $\lambda$6376 are seen.} 
   \label{fig:spec}
\end{figure}

Since colors and color changes can be used as a parameter to 
discriminate TDE from other types of transients/variations \citep{van+21},
we calculated the $zg-zr$ values during the flare time (right panel
of Figure~\ref{fig:lcfit}).
The two-band magnitudes obtained in the same night were considered to be
simultaneous and used in the calculation. 
Over $\sim$200\,days, the observed $zg-zr\sim 0.3$, but we should 
take into account the color in quiescence especially when the magnitude changes 
caused by the flare were not very large. As J1034 had the average 
$zg-zr\simeq 0.636\pm0.005$ in quiescence, $zg-zr$ values ($\sim -0.34$) 
shown in the right panel of Figure~\ref{fig:lcfit} were
calculated by subtracting the quiescent color value from the observed ones
(similar information also presented in Section~\ref{subsec:bb}). 
Only a small rate of $2.2\pm0.8\times 10^{-4}$\,day$^{-1}$ for the color
values during the flare was derived. 
Thus the blue color during the flare and its close-to-zero
rate are consistent with those of TDEs \citep{van+21}.
\begin{figure}
   \centering
   \includegraphics[width=\textwidth, angle=0]{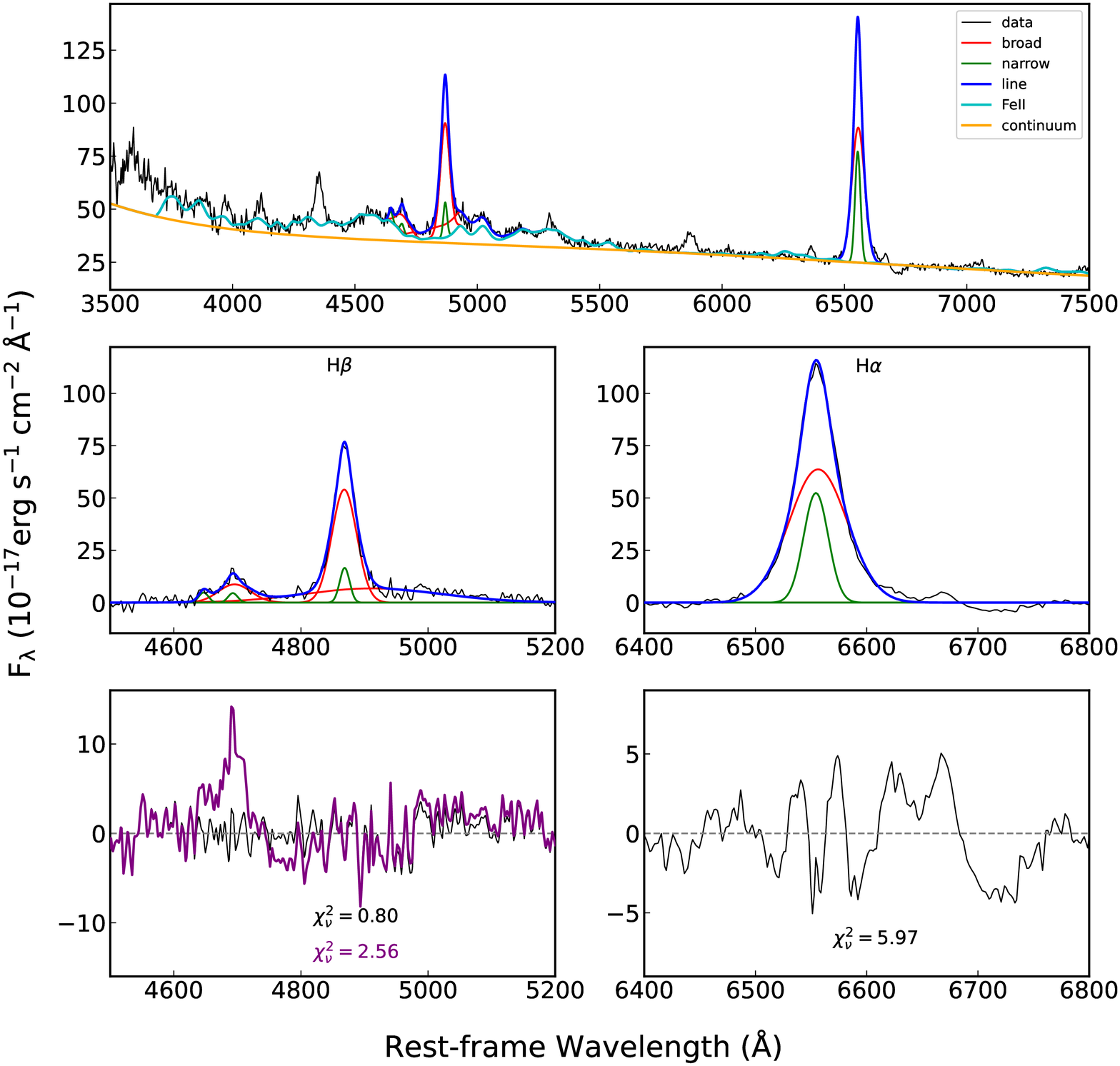}
	\caption{Example of the fitting to the optical spectra
	of J1034, where the first one (in Table~\ref{tab:obs}) is shown.
	The fitting included the Fe~II emission (3686--7484\,\AA; top panel).
	Two main regions covering H${\alpha}$ and H${\beta}$ are 
	shown in the middle and bottom panels (line fitting and residuals, 
	respectively).
	One broad (red line) and one narrow (green line) component
	are needed to fit the prominent H${\alpha}$ and H${\beta}$ lines.
	In the H${\beta}$ region, the existence of
	the lines He~II $\lambda$4686 and N~III $\lambda$4640 is needed; 
	otherwise a residual bump at their
	position would appear and induce larger $\chi^2$ in the fitting.} 
   \label{fig:specfit}
\end{figure}

\subsection{Features of the optical spectra}
\label{subsec:spec}

The obtained spectra are shown in Figure~\ref{fig:spec}. Broad hydrogen
Balmer (H${\alpha}$--H${\delta}$) and helium (He II $\lambda$4686 and 
He I $\lambda$5875) emission lines were detected. 
It can be noted that H${\alpha}$ on April 24 (the third one
in Table~\ref{tab:obs}) appears as the strongest.
In addition, there are
Fe II emission (e.g., \citealt{bla+17,he+21}) visible, particularly in 
the 5000--5500\,\AA\ region (see also Figure~\ref{fig:specfit}), and
weak but notable [Fe XIV] $\lambda$5304 and [Fe X] $\lambda$6376 emission
lines (e.g., \citealt{wan+12}). 
\begin{table}
   \bc
   \begin{minipage}[]{100mm}
   \caption[]{Measurements for the emission features\label{tab:line}}\end{minipage}
   \setlength{\tabcolsep}{7pt}
   \small
   \begin{tabular}{lccccc}
   \hline\noalign{\smallskip}
Line    & 2022-04-23            & 2022-04-23         & 2022-04-24           & 2022-04-25          & 2022-06-03\\
        & (First)               & (Second)           &                      &                     &           \\
   \hline\noalign{\smallskip}
	   H$\alpha$ broad & & & & &  \\
	   FWHM      & 2427.1$\pm$41.8       & 2501.1$\pm$16.6    & 2891.6$\pm$60.7      & 3048.4$\pm$97.1    & 2696.1$\pm$59.5\\
	   %%2257.7$\pm$91.5       & 2294.1$\pm$30.3    & 2515.5$\pm$261.7     & 2209.9$\pm$429.1   & 2515.6$\pm$104.1\\
           EW           & 199.7$\pm$10.6        & 201.5$\pm$3.2      & 185.9$\pm$10.5       & 195.1$\pm$16.3     & 184.5$\pm$10.2\\
	   %%226.1$\pm$9.1         & 228.2$\pm$7.9      & 206.1$\pm$17.9       & 245.3$\pm$29.5     & 188.6$\pm$8.3\\
           Flux       & 4956.6$\pm$262.1      & 4182.1$\pm$67.3    & 5305.1$\pm$300.6     & 4256.3$\pm$357.1   & 2439.0$\pm$134.9\\
	   %%5553.1$\pm$223.6      & 4722.9$\pm$162.5   & 5813.0$\pm$504.9     & 4746.6$\pm$571.2   & 2533.9$\pm$111.1\\\hline
	   H$\alpha$ narrow & & & & & \\
	   FWHM      & 850.2$\pm$28.5        & 845.9$\pm$1.5      & 848.7$\pm$15.3       & 848.7$\pm$7.8      & 825.5$\pm$17.5\\
	   %%917.7$\pm$151.2       & 917.6$\pm$151.1    & 807.7$\pm$80.7       & 837.6$\pm$93.9     & 770.9$\pm$58.1\\
	  EW       & 22.3$\pm$3.3          & 31.1$\pm$1.0       & 113.7$\pm$4.5        & 111.5$\pm$4.6      & 100.4$\pm$4.6\\
	   %%14.3$\pm$7.8          & 24.2$\pm$7.6       & 102.5$\pm$18.3       & 89.3$\pm$31.6      & 87.8$\pm$9.7\\
           Flux   & 552.9$\pm$80.7        & 645.9$\pm$21.7     & 3242.9$\pm$128.4     & 2431.2$\pm$100.4   & 1328.6$\pm$61.4\\
	   %%353.5$\pm$192.3       & 501.5$\pm$156.5    & 2887.9$\pm$515.0     & 1727.0$\pm$610.5   & 1180.2$\pm$130.7\\\hline
	   H${\beta}$ broad & & & & & \\
	   FWHM       & 2831.9$\pm$18.0       & 2882.3$\pm$48.9    & 2881.3$\pm$79.5      & 2860.2$\pm$75.9    & 3308.4$\pm$157.7\\
	   %%2997.1$\pm$110.2      & 3699.1$\pm$214.6   & 2379.3$\pm$202.4     & 3860.1$\pm$172.9   & 2922.3$\pm$293.3\\
           EW         & 76.8$\pm$1.6          & 77.2$\pm$2.9       & 55.8$\pm$5.0         & 57.9$\pm$5.8       & 59.6$\pm$7.7\\
	   %%135.1$\pm$2.7         & 160.9$\pm$4.3      & 66.9$\pm$6.9         & 85.5$\pm$4.9       & 69.3$\pm$6.6\\
           Flux          & 2641.9$\pm$56.7       & 2330.1$\pm$88.7    & 1367.7$\pm$123.1     & 1156.7$\pm$117.4   & 1057.6$\pm$138.2\\
	   %%4296.4$\pm$85.7       & 4455.1$\pm$119.3   & 1700.1$\pm$176.2     & 1498.8$\pm$86.7    & 1218.8$\pm$116.5\\\hline
	   H$\beta$ narrow & & & & & \\
	   FWHM         & 1123.8$\pm$12.6       & 1122.6$\pm$14.4    & 1116.1$\pm$20.3      & 1085.4$\pm$46.3    & 1096.3$\pm$44.9\\
	  %% 920.6$\pm$151.5       & 921.4$\pm$151.6    & 918.7$\pm$139.9      & 916.4$\pm$135.6    & 917.5$\pm$139.0\\
           EW           & 9.3$\pm$0.5           & 11.2$\pm$0.9       & 37.8$\pm$2.4         & 32.6$\pm$3.7       & 38.9$\pm$3.2\\
	   %%7.3$\pm$3.1           & 13.3$\pm$4.3       & 22.2$\pm$7.8         & 30.1$\pm$5.5       & 29.0$\pm$7.9\\
           Flux         & 320.6$\pm$17.8        & 338.6$\pm$28.8     & 926.6$\pm$60.5       & 650.4$\pm$74.4     & 691.1$\pm$58.1\\
	   %%230.1$\pm$96.5        & 364.3$\pm$118.2    & 562.8$\pm$198.6      & 526.6$\pm$95.7     & 509.7$\pm$138.9\\\hline
	   HeII broad & & & & & \\
	   FWHM         & 3648.0$\pm$90.0       & - & - & - & - \\
	   %%5797.5$\pm$357.6   & 7915.6$\pm$901.1     & 3955.4$\pm$447.4   & 5900.6$\pm$1228.9\\
	   EW  & 15.0$\pm$0.7    & - & - & - & - \\       
	   %%& 17.8$\pm$2.1       & 19.6$\pm$4.2         & 19.1$\pm$4.2       & 10.6$\pm$4.1\\
	   Flux & 530.5$\pm$24.9   & - & - & - & - \\     
	   %%& 557.0$\pm$66.5     & 484.4$\pm$104.4      & 389.6$\pm$86.2     & 203.5$\pm$79.3\\
	   HeII narrow & & & & & \\
	   FWHM & 1123.8$\pm$12.6       & 1122.6$\pm$14.4    & 1116.1$\pm$20.3      & 1085.4$\pm$46.3    & 1096.3$\pm$44.9\\
	   EW & 2.3$\pm$0.1           & 3.4$\pm$0.2        & 2.8$\pm$1.9          & 2.9$\pm$0.7        & 6.1$\pm$1.0\\
	   Flux & 80.5$\pm$3.4          & 106.02$\pm$7.7     & 57.3$\pm$52.9        & 59.9$\pm$16.1      & 117.9$\pm$19.3\\
	   Fe X & & & & & \\
	   FWHM & 1047.2$\pm$78.1       & 842.3$\pm$87.9     & 788.6$\pm$94.5       & 823.1$\pm$73.7     & 978.1$\pm$168.4\\
           EW & 4.7$\pm$0.7           & 3.9$\pm$0.7        & 9.0$\pm$2.1 .        & 12.9$\pm$2.2       & 5.9$\pm$2.1\\
	   Flux & 124.1$\pm$17.9        & 84.5$\pm$16.7      & 249.4$\pm$58.5       & 271.0$\pm$48.0     & 78.9$\pm$27.3\\
	   Fe XIV & & & & & \\
	   FWHM & 1019.8$\pm$120.9      & 1108.3$\pm$86.8    & 951.3$\pm$208.1      & -                  & -\\
	   EW & 3.9$\pm$0.8           & 4.0$\pm$0.7        & 5.6$\pm$2.1          & -                  & -\\
	   Flux & 124.4$\pm$26.4        & 111.0$\pm$19.8     & 137.6$\pm$52.1       & -                  & -\\
	   Fe II emission  & & & & & \\
	   Flux	   & 12224.7$\pm$3373.8    & 10554.7$\pm$228.9  & 11231.8$\pm$3558.6   & 8960.9$\pm$22.7   & 8090$\pm$86.6\\
	   %%13080.3$\pm$84.2      & 12668.2$\pm$26.2   & 10000.2$\pm$130.8    & 7620.9$\pm$147.6   & 8242.9$\pm$94.3\\
\noalign{\smallskip}\hline
\end{tabular}
\ec
	\tablecomments{0.86\textwidth}{Full width at Half Maximum (FWHM),
	equivalent Width (EW), and flux measurements are in units of 
	km\,s$^{-1}$, angstrom (\AA), and 10$^{-17}$\,erg\,cm$^{-2}$\,s$^{-1}$,
	respectively.}
\end{table}

In order to study the properties of these emission features,
we used the Python QSO fitting code (PyQSOFit; \citealt{gsw18}). As shown
in Section~\ref{subsec:src}, J1034 is likely an AGN when in its quiescent 
state.
Also in our fitting, we could only find a weak (or rather uncertain) galaxy
component. Thus the code likely provided sufficiently good results
for the properties of the emission features.
The main regions studied were those around the H${\alpha}$ and H${\beta}$ 
respectively, and the Fe II emission from 3686\,\AA\ to 
7484\,\AA\ \citep{bg92} was considered as well.

Since we found that a single Gaussian function could not well fit the two 
prominent Balmer lines, we added a narrow component.  
An example of the spectral fitting
is shown in Figure~\ref{fig:specfit}. 
In addition, the line He II $\lambda$4686 was fitted in the same 
way, as a broad component was needed in the fitting.
The lines [Fe XIV] $\lambda$5304 and [Fe X] $\lambda$6376 were fitted
with one component, although  
for the last two of our spectra, [Fe X] could not be determined.
The measurement results for the five spectra are given in Table~\ref{tab:line}.
The uncertainties of the line measurements were derived 
from a MCMC method recently included in the code, and for the Fe~II emission,
we followed the method in \citet{he+21} and the uncertainties
were estimated as the standard deviations of the results from fitting 
100 times.
Based on the measurements for the narrow components, redshift 
$z=0.136$ was determined.

In the fitting, we suspect that the two components, a broad plus a narrow one, 
may not be able to correctly derive the strengths of each of them,
although they can well describe the total line profiles.
For example, there is not a clear trend for the changes of the line components,
and the brightest H${\alpha}$ (on April 24) was caused by the enhancement
of the narrow component. There were also the cases that an additional
broad continuum-like component was needed in the fitting
(e.g., the H${\beta}$ region shown in
the left-middle panel of Figure~\ref{fig:specfit}).
In any case, the FWHMs of the narrow components of the lines are generally
consistent with each other, as the values are in a range of 
800--1100 km\,s$^{-1}$. The FWHMs of the broad components of H$\alpha$ and 
H$\beta$ can also be noted to mostly be $\sim 3000$~km\,s$^{-1}$.
In addition, the Fe II emission is shown to be roughly declining from 
the first to the last.

The presence of the
line He~II $\lambda$4686, as well as N~III $\lambda$ 4640, was required; 
otherwise the fitting would end with a residual bump at the position and 
a larger $\chi^2$ value (see Figure~\ref{fig:specfit}). However the broad
component in the fitting was highly uncertain, probably due to low
signal-to-noise ratios of the detections and the mixture with 
the Fe~II emission.  We carefully examined the
the region and for the spectrum 2--5, the component was more likely added
so that the resulting $\chi^2$ could be minimized. We thus only report
the results for spectrum 1, for which the fitting is shown in 
Figure~\ref{fig:specfit}.

\subsection{Broad-band spectrum}
\label{subsec:bb}

Provided with the X-ray flux upper limit and UV flux obtained in the 
{\it Swift} observation, we constructed the broad-band 
spectrum of the source in the flaring state on 2022 April 28.
The ZTF measurements on the date were 17.730$\pm0.031$ and 17.427$\pm0.025$\,mag
in $zg$ and $zr$ band respectively. The average
quiescent optical fluxes in the two bands were subtracted from them.
For the UV flux, we estimated a quiescent value by linear interpolation
from the two GALEX data points and the value was subtracted as well.
As the Galactic extinction towards 
the source is $E(B-V)\simeq 0.032$ \citep{sf11}, we used this value and
the extinction law given by \citet{ccm89} to deredden the UV and optical 
fluxes. 
Then the wavelengths of the X-ray, UV, and optical were shifted back to
the rest frame with $z=0.136$. The resulting spectrum is shown in 
Figure~\ref{fig:bb}.

Since emission of TDEs has a thermal origin \citep{gez21}, 
we estimated the properties of the blackbody emission of the flare event
by fitting the broad-band spectrum.
Because there are only three data points, no effort was made to search for
a complete range of blackbody parameters that can provide fits to
the data points. Instead, we tested different temperature values and searched
for the best fits around the values. We found
a likely good fit when the blackbody temperature is 1.48\,eV (or 17160\,K)
and the spherical radius is 2.9$\times 10^{-4}$\,pc 
(or 8.9$\times 10^{14}$\,cm for a source distance of 665\,Mpc). 
The reduced $\chi^2$ is large, $\chi^2\simeq 15$.
This large $\chi^2$ value probably reflects that the spectral data points
would have larger uncertainties; for example, the quiescent UV flux 
was estimated from the measurements obtained long time ago and
the times of the data points
were only on the same date but not exactly simultaneous.

\section{Discussion}
\label{sect:dis}

Exploring the ZTF data, we have found a TDE-like optical flaring event.
We have requested follow-up observations and conducted multi-wavelength
studies of the event source.
By studying the ZTF light curves, we have found that the flare can be 
described with a typical TDE model, a Gaussian rise
plus a power-law decline. While the light curve data and power-law index
do not follow the exact $\alpha=-5/3$ decline, there have been many
cases showing variations from the typical decline (e.g., \citealt{van+21}) and
the physical reasons for the variations have been discussed (e.g.,
\citealt{lr11,cn19}). From the fitting, we have estimated that the event 
started from MJD~59489 (or $-$33\,days from $t_p$) and would last for 
more than 500\,days. In addition, 
the optical color $zg-zr$ and its changes
during the flare were found to be $\sim -0.34$ and close to zero respectively.
These values
are typical for TDEs (e.g., \citealt{van+21}).

We have also found MIR brightening from the WISE light curve data. The flux 
increments
were 1.4 and 1.1\,mJy in W1 and W2 band respectively. The corresponding
luminosity of the flare at W1 band was 
$\nu L_{\nu} \simeq 7.4\times 10^{43}$\,erg\,s$^{-1}$. This value
is substantially higher than those derived for TDEs found in non-active
galaxies \citep{jia+21} but in the range for the TDEs in AGN \citep{van+21b}.
The luminosity difference has been discussed to reflect the available dust
covering the central MBHs; AGN contain a dust torus such that they are seen
to show more luminous MIR emission when TDEs occur \citep{jia+21}.
The high MIR luminosity thus supports J1034 as an AGN when the source is
in quiescence.
The average
time for the brightening is MJD~59545, which is larger than $t_p$ in our 
optical light curve fitting by 23\,days. Because we do not know the 
exact peak time in the MIR data due to the large observational gaps of 
the data and the MIR brightening in TDEs usually has a delay time, 
i.e., the MIR peak time with respect to $t_p$, in a range of 
$\sim$100--200\,days \citep{van+21b}, we may consider that the delay 
time $\tau$ would be $\tau \geq 23$\,day.
This delay time can be used to estimate the distance of the dust
to the MBH in this system, which would be
$\sim c\tau/(1+z)\geq 0.02$\,pc (where $c$ is the speed of light).
\begin{figure}
   \centering
   \includegraphics[width=0.5\textwidth, angle=0]{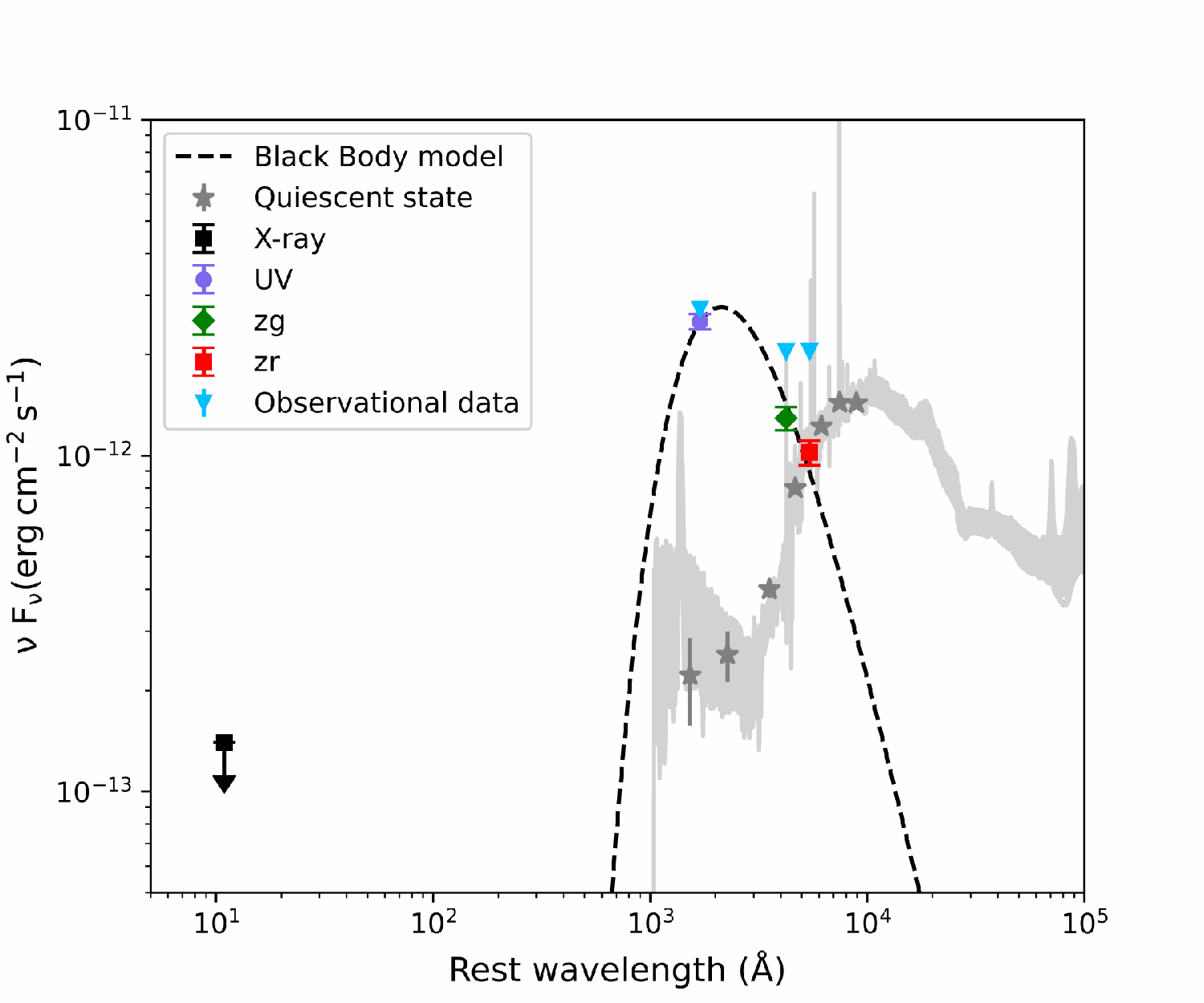}
	\caption{Optical and UV fluxes and X-ray flux upper limit
	of J1034 on 2022 April 28 (MJD~59697). The quiescent SED in
	Figure~\ref{fig:sed} is plotted to show the flux changes.
	After being subtracted with the corresponding quiescent fluxes, 
	the optical and UV data points can approximately be fitted with a 
	blackbody of temperature 
	1.48\,eV and spherical radius 2.9$\times 10^{-4}$\,pc (dashed curve).}
   \label{fig:bb}
\end{figure}

The optical spectra contain H and He emission lines, which would
suggest that this candidate TDE belongs to the H+He class. However given
the host likely being an AGN, the emission features or some of them could
be intrinsic ones. We note that our spectra are very similar to those
reported for ASASSN-18jd \citep{neu+20}. The both have 
FWHMs$\sim$3000\,km\,s$^{-1}$ broad H$\alpha$ and H$\beta$ lines, weak
He II $\lambda$4684 and N~III $\lambda$4640 lines that are overlapped with
the Fe II emission, and weak coronal lines [Fe XiV] and [Fe X]. \citet{neu+20}
have extensively discussed the features and their variations over 
the flare event,
and concluded that ASASSN-18jd could be a TDE or a new type of 
transient driven by the MBH. In any case, 
spectroscopy of J1034 in quiescence would help clarify the nature of 
AT2021acak. We note that in the TDE scenario, the Fe II emission is 
considered as
the result of sublimation of dust (e.g., \citealt{he+21}) 
by strong radiation of the TDE and the coronal lines
as the result of photoionization of gas (e.g., \citealt{wan+12,gez21}).

Using the UV flux measurement provided by the {\it Swift} observation,
we have estimated a blackbody temperature of $\sim$1.48\,eV and a radius of
$\sim 8.9\times 10^{14}$\,cm. This blackbody has negligible emission
at X-rays (Figure~\ref{fig:bb}), which made us not consider the marginal 
detection in 0.3--2\,keV with {\it Swift}/XRT.  However we note that 
ASASSN-18jd had X-ray emission whose luminosities were approximately an order 
of magnitude lower than those of its UV/optical
blackbody (see Figures 3 \& 5 in \citealt{neu+20}), and the X-ray emission
contained a soft, $\sim$0.1 keV blackbody component. Examining our
broad-band spectrum in Figure~\ref{fig:bb}, the X-ray flux (shown as an upper
limit), when we consider
the marginal detection, would appear to be similar to that of ASASSN-18jd
(i.e., the X-ray flux is an order of magnitude lower than that of the 1.48\,eV
blackbody).
Thus there could be soft X-ray emission from J1034 just like in ASASSN-18jd 
and the similarity between the two cases would be further established.

The luminosity of the UV/optical blackbody for J1034
can be estimated to be 4.9$\times 10^{43}$\,erg\,s$^{-1}$.
The temperature and luminosity put this source right into the group
of optically selected TDEs (see Figure~2 in \citealt{gez21}). As a case with
radius $<10^{15}$\,cm would generally indicate a H+He class 
TDE \citealt{gez21},
the radius of the blackbody in this respect is also consistent by showing 
both H plus He 
emission lines in the flare event of J1034. However we note that
the blackbody
luminosity is comparable to the MIR value, which suggests that the blackbody
emission would have dropped significantly from the peak to 2022 April 28 
(MJD~59697) over $\sim$175 days. Considering the ratios of 
the peak luminosities at the MIR to those of the blackbody
derived for TDEs \citep{jia+21,wan+22}, the peak blackbody
luminosity of this flare would have been at least 10 times that of the MIR, 
which is
$\sim 10^{45}$\,erg\,s$^{-1}$ and would imply that the blackbody temperature
(assuming a fixed radius) had a factor of 2 decline over the course. 
Even assuming such a decline, the estimated peak temperature and luminosity 
would still put this case into the same TDE group \citep{gez21}.

As a summary, the transient AT2021acak has been found to have the
following properties: 1) optical light curves can be described with a Gaussian
rise plus a power-law decay while with color $zg-zr\sim -0.34$
and close-to-zero color changes during the 
decay; 2) optical spectra show emission features that include the prominent 
Balmer lines, weak helium and coronal Fe lines, and Fe~II emission; 3) UV and
optical emission can be described with a $\sim$1.48\,eV blackbody with a 
spherical radius of $\sim 8.9\times 10^{14}$\,cm. These properties can be
indicative evidence of an H+He class TDE, while the limited observational
data we have and a likely AGN host complicate the identification. 
However as discussed in detail by \citet{neu+20} about the 
features of ASASSN-18jd, the similarity of AT2021acak to it
would suggest a now-so-called ambiguous 
nuclear transient (ANT; see, e.g., \citealt{tra+19,hin22}) as another 
possibility.
The number of these ANTs identified is rapidly increasing \citep{hin22}, and
their nature remains to be understood.
In any case, further 
observations of the event are warranted as the property changes in the decline
process would be obtained. Moreover, when the source goes back to its
quiescent state, optical spectroscopy will be required so that we would
have a better understanding of this transient event
by obtaining properties of the host galaxy. 

\begin{acknowledgements}
This work was based on observations obtained with the Samuel Oschin Telescope 
48-inch and the 60-inch Telescope at the Palomar Observatory as part of 
the Zwicky Transient Facility project. ZTF is supported by the National 
Science Foundation under Grant No. AST-2034437 and a collaboration including 
Caltech, IPAC, the Weizmann Institute for Science, the Oskar Klein Center 
at Stockholm University, the University of Maryland, Deutsches 
Elektronen-Synchrotron and Humboldt University, the TANGO Consortium of 
Taiwan, the University of Wisconsin at Milwaukee, Trinity College Dublin, 
Lawrence Livermore National Laboratories, and IN2P3, France. Operations are 
conducted by COO, IPAC, and UW.  
This work made use of data products from the Wide-field Infrared Survey 
Explorer, which is a joint project of the University of California, 
Los Angeles, and the Jet Propulsion Laboratory/California Institute of 
Technology, funded by the National Aeronautics and Space Administration.

We are grateful to {\it the Neil Gehrels Swift Observatory} for approving 
the requested ToO observation.
We acknowledge the support of the staff of the Lijiang 2.4-m telescope (LJT).  
Funding for the LJT has been provided by the CAS and the People’s Government 
of Yunnan Province. The LJT is jointly operated and administrated by YNAO and 
the Center for Astronomical Mega-Science, CAS.

We thank the anonymous referee for very helpful and detailed comments, 
	based on which this manuscript has been greatly improved.
	This research is supported by the National SKA program of China 
	(No. 2022SKA0130101), Basic Research Program of Yunnan Province 
	(No. 202201AS070005), and the National Natural Science Foundation of 
	China (12273033). Z.W. acknowledges the support by the Original
	Innovation 
	Program of the Chinese Academy of Sciences (E085021002).
\end{acknowledgements}

%%\appendix                  %%appendicial material is supported

%%\section{This shows the use of appendix} A postscript file is actually an ASCII text file (you may even edit it).  However, you need to transfer a PDF file or any compressed or packaged file in binary mode when using FTP.

%%\section{What is SCI?} SCI is the abbreviation of Science Citation Index system powered by the Institute for Scientific Information (ISI). For details please visit {\it http://apps.isiknowledge.com}.

\bibliographystyle{raa}
\bibliography{ms2022-0243}

\label{lastpage}

\end{document}